\documentclass{article}
\usepackage[T1]{fontenc}
\usepackage[utf8]{inputenc}
\usepackage{ismir} 
\usepackage{amsmath,cite,url}
\usepackage{graphicx}
\usepackage{color}
\usepackage{booktabs}
\usepackage{amsmath,amssymb}
\usepackage{booktabs}
\usepackage{array}
\usepackage{tabularx}
\usepackage{multirow}
\usepackage{enumitem}
\usepackage[bookmarks=false]{hyperref}
\usepackage{caption}
\usepackage[table]{xcolor}

\hypersetup{colorlinks=true,citecolor=blue,linkcolor=blue,urlcolor=blue}

\usepackage{siunitx}
\title{On the Geometry of Music Bandwidth Extension in Latent Spaces of Audio Codecs}

\multauthor
  {Hendrik Vincent Koops$^*$ \hspace{1cm} Hao Hao Tan$^*$ \hspace{1cm} Elio Quinton}
  {
  Music \& Audio Machine Learning Lab \\ 
  Universal Music Group, London, U.K.\\
  {\small \texttt{\{vincent.koops, harry.tan, elio.quinton\}@umusic.com} \hspace{0.5cm} $^*$Equal contribution  
  }
  }

\def\authorname{H. V. Koops, H. H. Tan, and E. Quinton}

\begin{document}

\maketitle

\begin{abstract}
Recent audio restoration increasingly relies on large-scale conditional latent generative modeling, including diffusion, Schr\"odinger Bridges, and Flow Matching variants, to invert degradations such as bandwidth limitation or noise.
We present an analysis of the performance of various state-of-the-art methods compared to simple arithmetic transformations in the latent spaces of multiple neural codecs for musical bandwidth extension. 
We show that estimating a single transport vector between the clean and degraded latent centroids on a reference set, and adding it to degraded latents, can yield restoration performance competitive with large diffusion models. 
This suggests, first, that some neural codec latent spaces exhibit structure aligned with audio bandwidth; and second, that in such cases complex conditional models may offer only limited gains over a simple vector addition.
We argue that these findings reveal an interesting avenue for future research whereby models could take advantage of the latent space structure in order to offer greater training and parameter efficiency, and overall better performance. 
Additionally, we propose to consider this simple arithmetic transformation as a baseline for music bandwidth extension research, as it allows an assessment of the contribution of learnable parameters towards restoration performance. 

\end{abstract}

\begin{figure}
    \centering
    \includegraphics[width=\columnwidth, trim={6.5cm 7cm 12cm 2cm},clip]{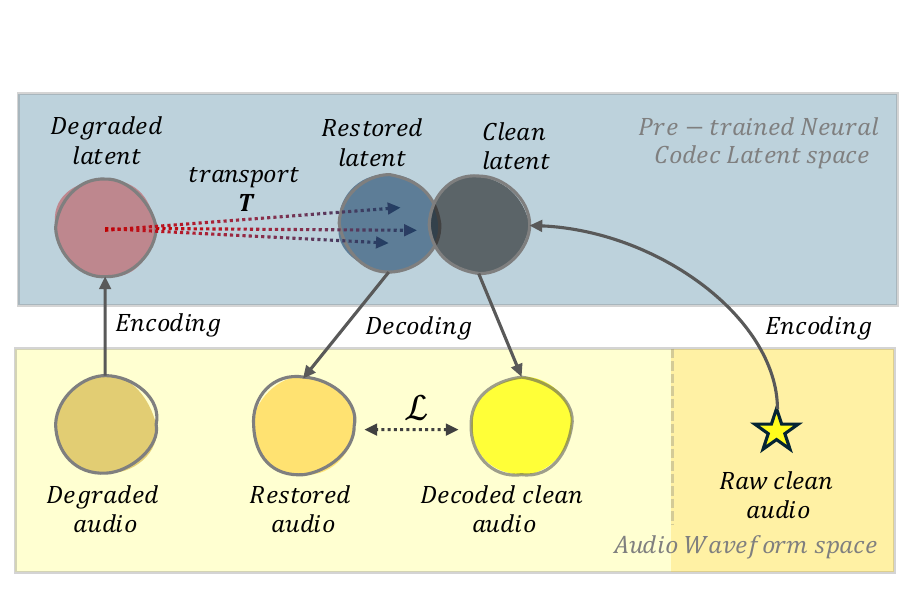}
\caption{We analyze audio restoration through latent transport geometry. For bandwidth extension, simple low-parameter latent transports (\textit{T}) already achieve strong waveform results on common audio metrics ($\mathcal{L}$), in some cases approaching large generative baselines. For other degradations, the latent correction is less coherent, showing that restoration difficulty depends strongly on the encoder and the degradation.}
    \label{fig:placeholder}
\end{figure}

\section{Introduction}

Many audio restoration tasks, such as bandwidth extension, require regenerating information and content that is not present in the original signal, which makes generative models an attractive approach for solving this type of problem.
Following the recent developments in generative modeling and the very significant improvement in the quality of generated content (audio, image, text, video etc) they brought \cite{esser2024scaling,labs2025flux,Evans2024LongFormMusic}, modern audio restoration increasingly relies on large generative models. 
Diffusion models, Schr\"odinger bridges, and related approaches have recently delivered strong results for super-resolution, enhancement, and bandwidth extension~\cite{kuleshov2017audio,liu2023audiosr,moliner2024babe,debortoli2021dsb,lipman2023flowmatching,zhang2025voicebridge}. 

Content restoration methods have shown progress in performance in various domains \cite{li2025diffusion,lemercier2025diffusion,dhyani2025high}. In the music domain, restoration methods tend to follow a similar trend of applying modern generative models, of increasingly larger scale \cite{kong2025a2sb}. This trajectory also results in progress in performance, but there remains room for improvement in order for these methods to become applicable in challenging professional scenarios. In this paper we will focus on generative modeling for musical audio restoration. 

While some generative restoration methods operate on the content space directly \cite{kong2025a2sb,zhu2024musichifi}, many of these systems operate on latent spaces of neural codecs \cite{bralios2025latent,liu2023audiosr,zhang2025voicebridge}. Generative modeling on the latent domain is more computationally practical, but implicitly makes the structure of the latent space a critical component of the system. So far, however, relatively little attention has been given in the literature to studies of the latent space structure and its suitability for audio restoration tasks, or the impact of the neural encoder in the process.  

This paper proposes an analysis of some properties of the latent space of several open neural audio encoders for audio restoration: Stable Audio Open \textsc{vae}~\cite{evans2024stableaudioopen}, \textsc{Co\-di\-Co\-dec}~\cite{pasini2025codicodec}, \textsc{dac}~\cite{kumar2023dac}, and \textsc{Encodec}~\cite{defossez2022encodec}. Surprisingly, we find that calculating a single mean transport vector between the clean and degraded latent centroids on a reference set, and adding it to degraded latents, can yield restoration performance competitive with large diffusion models. 
This suggests, first, that some neural codec latent spaces exhibit structure aligned with audio bandwidth; and second, that in such cases complex and resource-intensive generative models may offer only limited gains over a simple vector addition.
We argue that these findings reveal an interesting avenue for future research whereby models could take advantage of the latent space structure in order to offer greater training and parameter efficiency, and overall better performance. 
Additionally, we propose to consider this simple arithmetic transformation as a baseline for music bandwidth extension research, as it allows an assessment of the contribution of learnable parameters towards restoration performance. 

We encourage readers to listen to the audio samples and compare restoration results between simple latent transport and learned models on our companion website: \url{https://ismir26latentgeo.github.io}.

\section{Related Work}


Recent audio restoration systems are increasingly built as large generative models. Early neural bandwidth-extension work already treated super-resolution as a learned conditional mapping~\cite{kuleshov2017audio,su2021bandwidth}, and more recent approaches have pushed this paradigm further with large-scale diffusion models for audio super-resolution and blind restoration~\cite{liu2023audiosr,moliner2024babe}. Schr\"odinger bridge formulations~\cite{kong2025a2sb, debortoli2021dsb,zhang2025voicebridge} and flow-matching methods~\cite{yun2025flowhigh,lipman2023flowmatching} provide related continuous-transport views of conditional generation. Together, these lines of work have increasingly tied restoration performance to the generative modeling method.

\subsection{Restoration in latent representations}

Restoration is also increasingly performed in learned latent spaces rather than directly in the content space (e.g. waveform or spectrogram) \cite{liu2023audiosr,zhang2025voicebridge}. Latent-domain bandwidth extension with \textsc{vae}s~\cite{bachhav2019abe} and more recent latent upsampling and upmixing methods~\cite{bralios2025latent} show the practical advantages of operating on compact encoded representations. In parallel, neural audio codecs such as \textsc{SoundStream}~\cite{zeghidour2021soundstream}, \textsc{Encodec}~\cite{defossez2022encodec}, \textsc{dac}~\cite{kumar2023dac}, \textsc{Music2Latent}~\cite{pasini2024music2latent}, and \textsc{CoDiCodec}~\cite{pasini2025codicodec} continue to be developed. Most prior work in this area focuses on improving the restoration model that operates on top of such learned representations. Comparatively little effort has gone into analyzing the structure induced by the representations themselves and, more generally, the impact of the neural codec. 

\subsection{Perceptual and geometric structure in audio latent spaces}

Because neural codecs are a critical component of latent generative systems, it is reasonable to expect that the geometry and overall structure of the latent space may influence the restoration process. In audio, VAEs have been shown to organize latent space in ways that reflect perceptual timbre relationships~\cite{esling2018generative}, and \textsc{vae}-based speech models can expose interpretable generative factors such as source-filter structure~\cite{sadok2023sourcefilter}.
Beyond timbre-space regularization, prior work has shown that musical audio latent spaces can support disentangled representations of timbre and pitch and meaningful interpolation in timbre space~\cite{luo2019learning,tatar2021latent}.
Similar observations in vision suggest that generative latent spaces can also contain meaningful near-linear directions associated with semantic transformations~\cite{harkonen2020ganspace,shen2020interfacegan}. One major challenge in audio generative modeling within the latent space is learning representations useful for generation, which requires capturing semantic content, while the primary training objective of most audio codecs is reconstruction quality \cite{ye2025codec}.
These observations motivate our central question: if audio encoders already organize musically or perceptually meaningful variation in latent space, do they also exhibit structure that could be relevant and/or exploited for audio restoration?




\section{Experiments}

For brevity, our study focuses on bandwidth extension (BWE). We wish to evaluate whether audio bandwidth is somehow represented in the geometry of the latent space of various neural codecs. For that we propose to use simple algebraic operations in the latent space. 
Let $z^{\mathrm{clean}}_{i}$ and $z^{\mathrm{deg}}_{i}$ denote the clean and degraded latents of a paired audio example $x_{i}^{\mathrm{clean}}$ and $x_{i}^{\mathrm{deg}}$. We study restoration through a low-complexity latent transformation of the form $\hat{z_i} = z_{i}^{\mathrm{deg}} + T$, where $\hat{z_i}$ is the restored latent, and $T$ is a simple deterministic transport rule estimated from artificially generated paired clean/degraded examples. 

\subsection{Datasets and codecs}

To cover a wide range of complementary musical styles, while allowing reproducibility, we choose the following open datasets: 
\textsc{mtd}, a dataset of classical musical themes
~\cite{ZalkowBAM20_MTD_TISMIR}; 
\textsc{ccmixter}, a dataset of stereo remixes with various genres
~\cite{ccmixter}; and 
\textsc{maestro}, a dataset of piano performances
~\cite{hawthorne2019enabling}. For \textsc{mtd} and \textsc{ccmixter}, we use an 80/20 train/test split, and for \textsc{maestro} we follow the official train-test split.
Note that in this work, we use the train splits to derive a simple vector displacement, but we do not \textit{train} a generative model. We keep the common \textit{train/test} nomenclature for simplicity.

To obtain degraded audio samples, we resample the ground truth audio to twice the cutoff frequencies (8, 16 and 24 {\unit{\kilo\hertz}}) similar to Kong et al. in~\cite{kong2025a2sb}.
Resampling back to 44.1 kHz gives us hard band-limited audio with cutoffs at 4, 8 and 12 {\unit{\kilo\hertz}}. Then, the paired degraded-clean samples are passed through a neural codec to obtain $z^{\mathrm{clean}}$ and $z^{\mathrm{deg}}$. We compare across four open codecs  widely used in previous work: Stable Audio Open \textsc{vae}~\cite{evans2024stableaudioopen}, \textsc{Co\-di\-Co\-dec}~\cite{pasini2025codicodec}, \textsc{dac}~\cite{kumar2023dac}, and \textsc{Encodec}~\cite{defossez2022encodec}. For all codecs, we use continuous latent embeddings, typically taking the latent vectors before the quantization bottleneck, if any.

\subsection{Latent Transport: Mean Shift}
\label{subsec:mean_shift_local_mean}

We explore latent-based bandwidth extension through a simple latent transport $T$.
Our probe is a single global transport, which we call \textit{mean shift}. Let $\Delta(z_i) = z_i^{\mathrm{clean}} - z_i^{\mathrm{deg}}$ denote the restoration displacement given a sample $i$. 
Given the train-split of a dataset, we randomly sample up to $N=8192$ chunks of paired clean and degraded 1.5 seconds audio samples. 
We derive the mean-shift vector $T$, which is the average displacement in latent space over the $N$ paired samples, $T = \frac{1}{N}\sum_{i=1}^{N}\Delta(z_i)$. 
We apply this restoration $T$ to every degraded latent $z_i^{\mathrm{deg}}$ in the test set using the same mean-shift vector. Intuitively, one can interpret this vector $T$ as the average linear direction that represents bandwidth extension in the latent space.

\begin{table}[t]
\centering
\setlength{\tabcolsep}{5pt}
\renewcommand{\arraystretch}{1.08}
\resizebox{\columnwidth}{!}{%
\begin{tabular}{llrrrrrr}
\multicolumn{2}{c}{}
& \multicolumn{2}{c}{4\,kHz}
& \multicolumn{2}{c}{8\,kHz}
& \multicolumn{2}{c}{12\,kHz} \\
\cmidrule(lr){3-4}\cmidrule(lr){5-6}\cmidrule(lr){7-8}
Method & Variant
& \textsc{lsd} $\downarrow$ & SiSpec $\uparrow$
& \textsc{lsd} $\downarrow$ & SiSpec $\uparrow$
& \textsc{lsd} $\downarrow$ & SiSpec $\uparrow$ \\
\midrule
Degraded roundtrip &  & 1.154 & 31.49 & 1.137 & 32.99 & 1.129 & 33.33 \\
\midrule
CQTDiff{$\dag$} &  & 1.154 & 31.49 & 1.137 & 32.99 & 1.129 & 33.33 \\
\textsc{ibar}{$\dag$} &  & \textbf{0.769} & 12.69 & 0.688 & 12.22 & 0.616 & 13.48 \\
A2SB{$\dag$} & 4-part$^\star$ & 0.773 & \textbf{34.32} & \textbf{0.659} & \textbf{41.69} & \textbf{0.545} & \textbf{42.60} \\
\midrule
VAE & mean shift & 1.185 & 11.34 & 1.164 & 11.34 & 1.130 & 11.36 \\
CodiCodec & mean shift & 0.975 & 7.94 & 0.938 & 7.97 & 0.923 & 7.91 \\
DAC & mean shift & 1.041 & 15.35 & 1.003 & 15.64 & 0.998 & 15.72 \\
Encodec & mean shift & 0.901 & 19.50 & 0.799 & 19.91 & 0.742 & 20.05 \\
\end{tabular}%
}
\caption{Bandwidth extension results on \textsc{maestro}. $\dag$: results as reported by Kong et al. in~\cite{kong2025a2sb}. $\star$ = 4-partitioning. Note: \textsc{VisQOL} is not reported for \textsc{maestro} in~\cite{kong2025a2sb}.}
\label{tab:maestro_bwe_baselines}
\end{table}

\subsection{Evaluation and Geometry Probes}
\label{subsec:eval_geometry_probes}

We evaluate restoration performance with objective metrics commonly used in audio restoration and enhancement research. 
For BWE, we report Log-Spectral Distance (\textsc{lsd})~\cite{erell1990estimation}, SiSpec~\cite{liu2021voicefixer}, and ViSQOL~\cite{chinen2020visqol}, which together capture spectral fidelity, spectral consistency, and perceptual quality. Beyond these perceptual metrics on audio, we further study the latent space geometry which contributes to the effectiveness of the restoration transport $T$. We measure the alignment of individual restoration vectors $\Delta(z_i)$ with the global mean-shift direction $T$ by calculating the average pairwise cosine similarity: 

\begin{equation}
\label{eq:cosine}
\cos(\theta) = \frac{1}{N} \sum_{i=1}^{N} \frac{\Delta(z_i) \cdot T}{\|\Delta(z_i)\| \, \|T\|}
\end{equation}

We also compare computed transport vectors and their alignment across datasets, cutoffs and tasks to explore the determining factors in the latent geometry. We further stress-test the number of samples needed to derive a transport $T$ that performs decent restoration. 
Together, these analyses show whether a degradation induces a coherent shared restoration direction in the latent space, whether that direction is stable across conditions, and whether the representation preserves sample identity.

For context, we compare the results of our experiments with several high-capacity baselines that perform at or close to the state of the art. AudioSR~\cite{liu2023audiosr} is a large-scale diffusion model for audio super-resolution; CQTDiff is a diffusion baseline operating on a constant-Q representation; \textsc{ibar}~\cite{kong2025a2sb,lee2025etta,wang2023audit} is a restoration model guided by natural language instructions; and A2SB~\cite{kong2025a2sb} is an audio-to-audio Schr\"odinger bridge model, which we report in no-partitioning, 2-partitioning, and 4-partitioning variants. 
Partitioning refers to an ensemble (of size 1, 2 or 4) of expert diffusion denoisers, where each denoiser is specialized to a part (e.g. half in the case of 2-partitioning) of the diffusion process~\cite{balaji2022eDiff-I}. 

\section{Results}

\begin{table*}[t]
\centering
\scriptsize
\setlength{\tabcolsep}{6pt}
\renewcommand{\arraystretch}{1.0}

\resizebox{0.87\textwidth}{!}{%
\begin{tabular}{llc S S S S S S S S S}
& & & \multicolumn{3}{c}{4 kHz} & \multicolumn{3}{c}{8 kHz} & \multicolumn{3}{c}{12 kHz} \\
\cmidrule(lr){4-6}\cmidrule(lr){7-9}\cmidrule(lr){10-12}
Method & Variant & Parameters
& {LSD$\downarrow$} & {SiSpec$\uparrow$} & {ViSQOL$\uparrow$}
& {LSD$\downarrow$} & {SiSpec$\uparrow$} & {ViSQOL$\uparrow$}
& {LSD$\downarrow$} & {SiSpec$\uparrow$} & {ViSQOL$\uparrow$} \\
\midrule

Degraded roundtrip & & -
& 1.75 & 21.74 & 3.39 & 1.81 & 27.26 & 3.23 & 1.85 & 28.97 & 3.15 \\
\midrule
AudioSR$^\dag$ & & $\pm$280M
& 1.75 & 21.74 & 3.39 & 1.81 & 27.26 & 3.23 & 1.85 & 28.97 & 3.15 \\

CQTDiff$^\dag$ & & $\pm$15M
& 1.74 & 10.62 & 1.75 & 1.63 & 17.42 & 1.78 & 1.57 & 21.62 & 2.00 \\

IBAR$^\dag$ & & $\pm$1B
& \bfseries 1.12 & 12.31 & 2.99 & \bfseries 0.92 & 12.94 & 3.52 & \bfseries 0.85 & 13.08 & 3.84 \\

A2SB$^\dag$ & no partitioning & $\pm$565M
& 1.33 & 25.51 & 2.55 & 1.05 & 33.10 & 3.20 & 0.87 & 35.34 & 3.93 \\

& 2-partitioning & $\pm$565M
& 1.29 & \bfseries 28.15 & 3.10 & 1.07 & \bfseries 34.36 & 3.71 & 0.88 & 35.97 & 4.20 \\

& 4-partitioning & $\pm$565M
& 1.77 & 27.56 & 3.44 & 1.59 & 34.25 & 3.82 & 1.51 & \bfseries 36.07 & \bfseries 4.27 \\

\midrule

VAE & mean shift & 0
& 1.29 & 10.01 & \bfseries 3.45 & 1.15 & 10.13 & \bfseries 3.84 & 1.14 & 10.13 & 3.78 \\

CodiCodec & mean shift & 0
& 1.17 & 7.03 & 3.22 & 1.08 & 7.09 & 3.41 & 1.05 & 7.08 & 3.59 \\

\textsc{dac} & mean shift & 0
& 1.23 & 13.37 & 3.14 & 1.15 & 13.98 & 3.54 & 1.08 & 14.10 & 3.85 \\

Encodec & mean shift & 0
& 1.64 & 8.10 & 2.31 & 1.59 & 11.07 & 3.18 & 1.55 & 11.21 & 3.57 \\

\end{tabular}%
}
\caption{Bandwidth extension results on \textsc{mtd}. $\dag$: results as reported by Kong et al. in~\cite{kong2025a2sb}.}
\label{tab:bwe_results_mtd}
\end{table*}

\begin{table*}[t]
\centering
\scriptsize
\setlength{\tabcolsep}{6pt}
\renewcommand{\arraystretch}{1.0}

\resizebox{0.87\textwidth}{!}{%
\begin{tabular}{llc S S S S S S S S S}
& & & \multicolumn{3}{c}{4 kHz} & \multicolumn{3}{c}{8 kHz} & \multicolumn{3}{c}{12 kHz} \\
\cmidrule(lr){4-6}\cmidrule(lr){7-9}\cmidrule(lr){10-12}
Method & Variant & Parameters
& {LSD$\downarrow$} & {SiSpec$\uparrow$} & {ViSQOL$\uparrow$}
& {LSD$\downarrow$} & {SiSpec$\uparrow$} & {ViSQOL$\uparrow$}
& {LSD$\downarrow$} & {SiSpec$\uparrow$} & {ViSQOL$\uparrow$} \\
\midrule

Degraded roundtrip & & -
& 2.00 & 12.50 & 2.74 & 1.86 & 14.93 & 3.09 & 1.75 & 18.35 & 3.51 \\
\midrule

AudioSR$^\dag$ & & $\pm$280M
& 2.00 & 12.50 & 2.74 & 1.86 & 14.93 & 3.09 & 1.75 & 18.35 & 3.51 \\

CQTDiff$^\dag$ & & $\pm$15M
& 2.01 & 14.67 & 1.97 & 2.06 & 15.88 & 1.86 & 2.10 & 16.34 & 1.85 \\

IBAR$^\dag$ & & $\pm$1B
& 1.64 & 7.11 & 2.37 & \bfseries 1.41 & 10.46 & 2.60 & 1.36 & 7.86 & 2.74 \\

A2SB$^\dag$ & no partitioning & $\pm$565M
& 1.93 & 14.05 & 2.77 & 1.71 & 19.95 & 3.20 & 1.48 & 27.17 & 4.04 \\

& 2-partitioning & $\pm$565M
& 1.85 & 18.00 & \bfseries 2.85 & 1.62 & \bfseries 23.39 & \bfseries 3.43 & 1.45 & \bfseries 29.26 & 4.21 \\

& 4-partitioning & $\pm$565M
& 1.84 & 17.46 & 2.65 & 1.65 & 23.17 & \bfseries 3.43 & 1.50 & 29.20 & \bfseries 4.23 \\

\midrule

VAE & mean shift & 0
& 1.57 & 10.21 & 2.74 & 1.44 & 10.67 & 3.12 & 1.31 & 10.74 & 3.63 \\

CodiCodec & mean shift & 0
& \bfseries 1.52 & 6.02 & 2.67 & \bfseries 1.41 & 6.33 & 2.96 & \bfseries 1.30 & 6.35 & 3.42 \\

\textsc{dac} & mean shift & 0
& 1.71 & 12.03 & 2.65 & 1.56 & 12.88 & 2.90 & 1.51 & 13.45 & 3.58 \\

Encodec & mean shift & 0
& 1.92 & \bfseries 18.22 & 2.72 & 1.83 & 18.72 & 2.93 & 1.73 & 18.93 & 3.49 \\

\end{tabular}%
}
\caption{Bandwidth extension results on \textsc{ccmixter}. $\dag$: results as reported by Kong et al. in~\cite{kong2025a2sb}.}
\label{tab:bwe_results_ccmixter}
\end{table*}

\subsection{Zero-parameter Bandwidth Extension}
\label{subsec:bwe_latent}

Tables~\ref{tab:maestro_bwe_baselines},~\ref{tab:bwe_results_mtd} and~\ref{tab:bwe_results_ccmixter} compare zero-parameter latent transports with high-capacity generative models for \textsc{bwe}.
Although mean shift is not intended as a practical replacement for these models, we were surprised to find that the performance gap between large models and a simple latent translation is much smaller than expected.
In fact the mean shift transport yields competitive performance on some metrics and some datasets. 
On \textsc{mtd} at 4\,kHz, for example, applying a mean-shift transport on \textsc{vae}, \textsc{Co\-di\-Co\-dec}, and \textsc{dac} latents yields lower \textsc{lsd} (1.17--1.29), and comparable or higher \textsc{ViSQOL} (3.14--3.45) than AudioSR ($\pm$280M parameters; \textsc{lsd} 1.75, ViSQOL 3.39) and A2SB without partitioning ($\pm$565M parameters; \textsc{lsd} 1.33, ViSQOL 2.55). On \textsc{ccmixter}, \textsc{vae} and \textsc{dac} consistently outperform \textsc{ibar}, a model with $\pm$1B parameters, with better SiSpec and ViSQOL values across all three cutoff frequencies. While high-capacity models remain stronger on some metrics, especially SiSpec, and on datasets like \textsc{maestro}, we argue that the performance gain is small in comparison with the difference in model complexity.

This suggests that current large restoration models may not yet be realizing their full potential. 
It also indicates on one hand that modification of the bandwidth is represented as a simple algebraic operation in the latent space, at least to some degree, and on the other hand that large generative models may not be taking full advantage of that latent space structure. 
We hope that this observation may open new research direction for developing possibly more efficient BWE models. For example, these models could attempt to leverage this latent space structure, so that most of the model capacity could be directed towards what the global mean shift cannot achieve, such as recovering sample-specific high-frequency detail, modeling the non-linear part of the transformation, or compensating for codec-specific decoder artifacts. 

As an additional benefit, our analysis reveals that different codecs construct latent spaces with different properties. Interestingly, we observe consistent patterns based on their metrics across different datasets. For example, CodiCodec gives the lowest \textsc{lsd}, but its SiSpec scores also remain low, which suggests that while it captures the broad spectral distribution, the reconstruction often misses spectral details that could distort the identity of the audio sample, which is detrimental to restoration performance. Encodec shows the opposite tendency in general, with high SiSpec but weaker \textsc{lsd}. Both \textsc{vae} and \textsc{dac} have a more balanced performance, with \textsc{dac} typically achieving a slightly better perceptual quality than \textsc{vae} in terms of SiSpec and ViSQOL. 
Taken together, we argue that opportunity for future improvement in music audio restoration may be found both by choosing or crafting a neural codec with adequate properties, and designing generative models that leverage the latent structure efficiently for the task.

\subsection{Cross-Dataset Consistency of Latent BWE Transport}
\label{subsec:cross-dataset}

We next ask whether the dominant BWE transport direction is tied to the musical dataset and degradation parameters or instead reflects the degradation itself more generally.
For each codec, dataset, and cutoff frequency, we compute the global mean-shift transport vector $T$ and compare the restoration vectors $\Delta(z_i)$ for all pairs of clean-degraded latents using cosine similarity using Equation \ref{eq:cosine}. 

\begin{figure*}[t]
    \centering
    \includegraphics[trim={1.35cm 2.5cm 1.3cm 0.7cm},clip,width=0.98\textwidth]{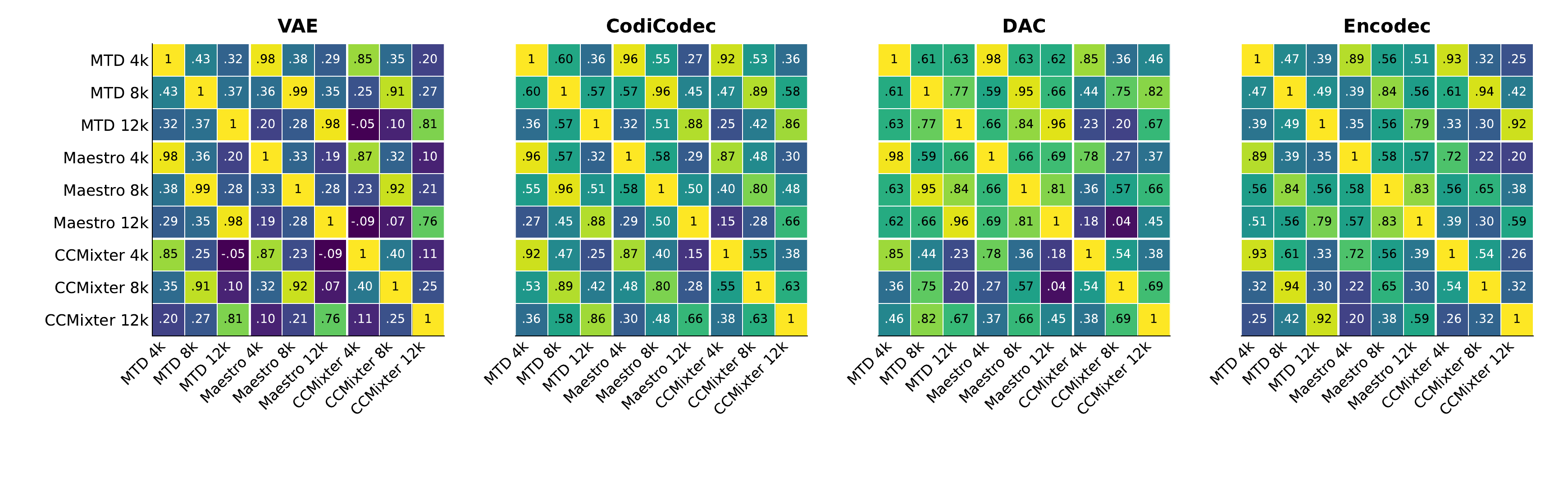}
    \caption{Per-codec cosine-similarity heatmaps of global mean-shift transport vectors across all dataset-cutoff conditions for bandwidth extension. Each panel corresponds to one codec and compares the nine conditions formed by the three datasets (\textsc{mtd}, \textsc{maestro}, \textsc{ccmixter}) and three cutoff frequencies (4, 8, and 12\,kHz). Brighter blocks indicate stronger alignment between learned transport directions.}
    \label{fig:transport_cosines_by_codec_heatmaps}
\end{figure*}

Figure~\ref{fig:transport_cosines_by_codec_heatmaps} shows that the dominant BWE transport is governed more by cutoff frequency than by dataset. For example in the \textsc{vae} case, cosine similarity calculated at the same cutoff frequency between the \textsc{mtd} and \textsc{maestro} mean-shift vectors reaches 0.98 at 4, 8, and 12\,kHz, and remains high at 0.85 between \textsc{mtd} and \textsc{ccmixter}. Conversely, within-dataset cross-cutoff similarities are much lower (0.43/0.32/0.37 on \textsc{mtd} and 0.33/0.19/0.28 on \textsc{maestro}).
This indicates that the dominant transport direction appears to be strongly determined by cutoff frequency - i.e. by the degradation transformation itself rather than by the dataset. This means that the same low-complexity restoration mechanism (embodied by vector $T$) transfers relatively well across musical domains.
This pattern is strongest for \textsc{vae} and \textsc{CodiCodec}, while \textsc{Encodec} and \textsc{dac} exhibit somewhat weaker and less stable alignment.

\begin{figure}[t]
    \centering
    \includegraphics[width=\columnwidth]{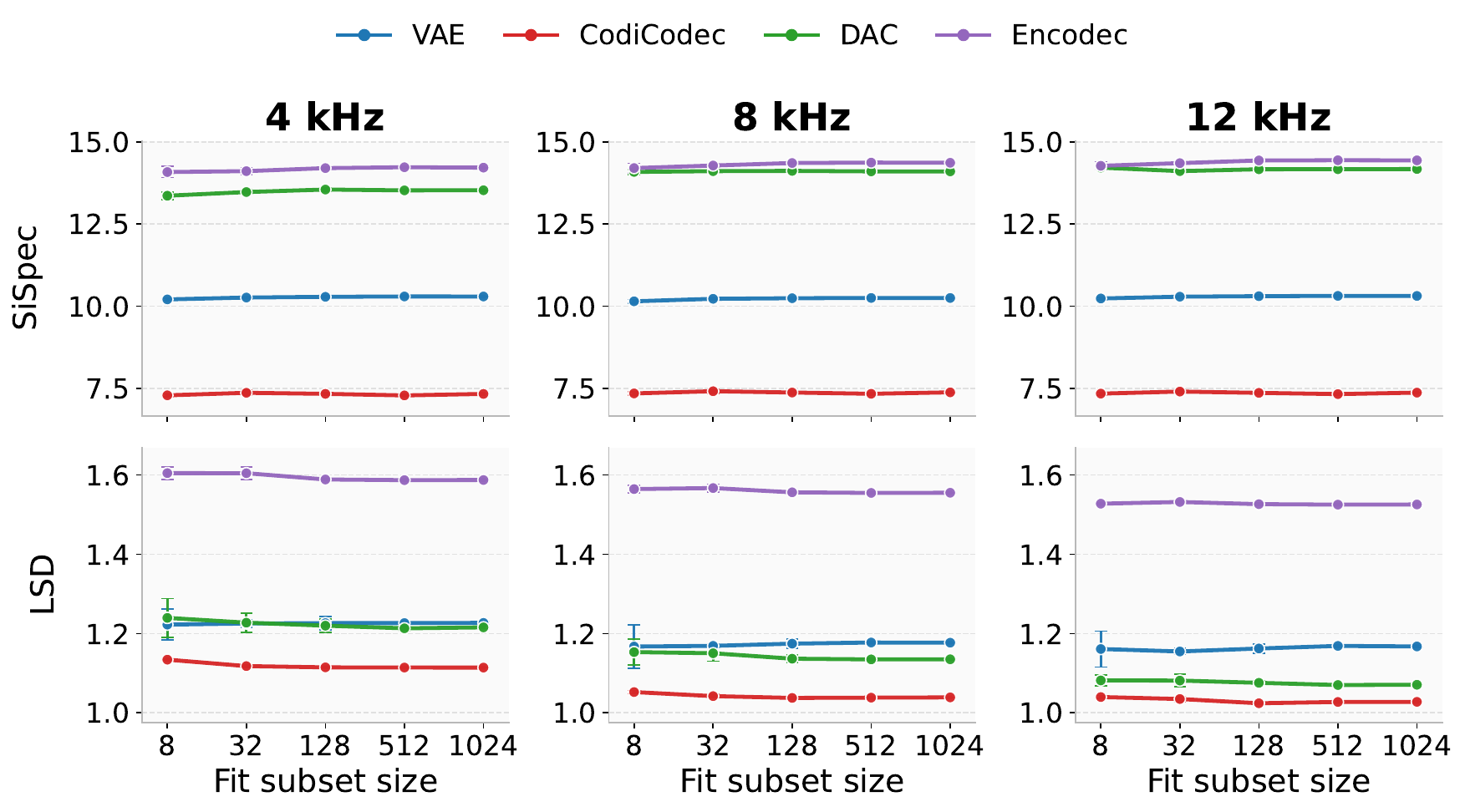}
    \caption{SiSpec and \textsc{lsd} as a function of fit subset size, grouped by cutoff and codec. Starting with just 8 samples, most metrics stay similar as fit subset size grows. Results are close to full-dataset results at 8 samples (as reported in Tables~\ref{tab:maestro_bwe_baselines}--\ref{tab:bwe_results_ccmixter}), suggesting strong sample efficiency for deriving effective transport probes.}
    \label{fig:sample_efficiency_decoded_sispec_raw_lsd}
\end{figure}

\subsection{Sample Efficiency of Mean-Shift Estimation Across Codecs}
\label{subsec:sample_efficiency_codecs}

Having found that \textsc{bwe} transport directions are often stable across datasets, we next ask how many paired examples are needed to estimate such a direction, as a proxy to measure how salient this direction is in the latent space. The fewer examples needed, the more salient. For each codec and cutoff, we calculate the global mean-shift vector $T$ across a range of random subset sizes and compare the resulting transport to the full-data estimate in both latent-space alignment and audio-space restoration metrics.

Figure~\ref{fig:sample_efficiency_decoded_sispec_raw_lsd} shows that for all codecs, SiSpec and \textsc{lsd} are nearly identical as the fit subset grows from 8 to 1024 examples. Thus, even a very small fit subset of 8 samples is sufficient to capture the global bandwidth transport vector for these probes, and that increasing the sample size further brings no improvement. This result suggests that the restoration direction is very salient in the latent space, therefore easy to estimate. 
Besides, the differences between codecs and cutoffs are much larger than the differences caused by fit-set size: Encodec remains strongest in SiSpec, while CodiCodec and \textsc{dac} generally maintain lower \textsc{lsd}. 



The takeaway from this result for future work on audio restoration is that estimating a rough restoration transport vector seems to be relatively easy for a given bandwidth degradation, and therefore that the modeling power of large generative models may unlock the next frontier in performance by modeling aspects of the transformation not captured by this salient direction, such as: sample-specific context and non-linear parts of the transformation.

\begin{figure}[t]
    \centering   \includegraphics[width=\columnwidth,trim={0cm 2cm 0cm 1.7cm},clip]{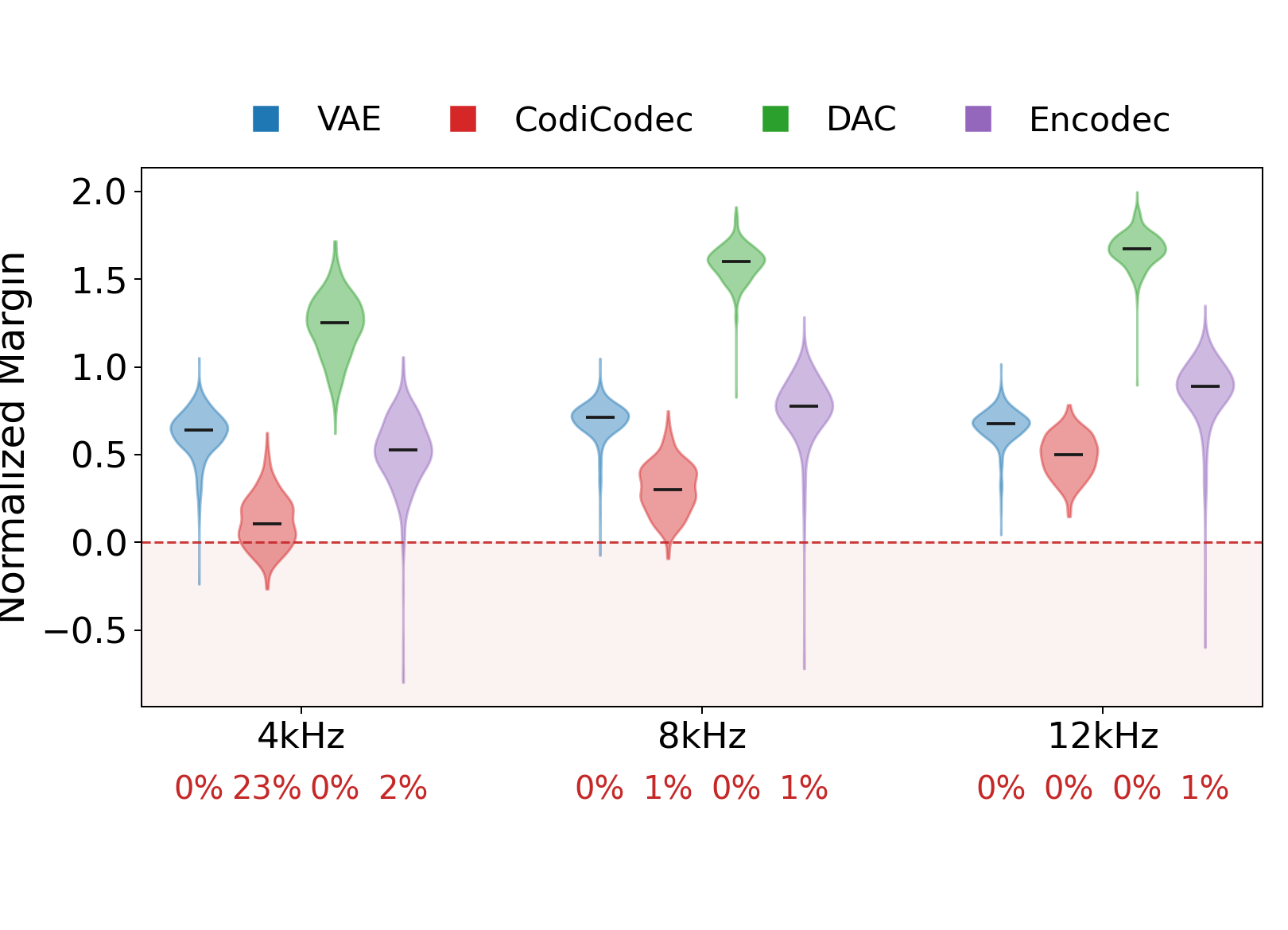}
    \vspace{-10px}
    \caption{Normalized BWE margin violin plots across codecs and cutoffs. Percentages beneath each codec indicate the share of samples with negative normalized margin. Lower percentage means better identity preservation.}
    \label{fig:structure_preservation_margin_orchestral}
\end{figure}

\subsection{Identity Preservation Analysis Across Codecs}
\label{subsec:identity}

We next examine how clean and degraded latents are distributed in the latent spaces. In particular, we explore whether the latent spaces exhibit a structure where pairs of clean and degraded latents tend to remain closer together than they would be to other samples in the latent space, or conversely if clean and degraded latents tend to be separated as much as two clean latents would be.
To that end, we define a margin $m = d_{\mathrm{inter}}^{\min} - d_{\mathrm{intra}}$ to measure the gap between intra-sample degraded-clean latent distance and the nearest inter-sample clean-clean latent distance. Positive values indicate that the clean latent remains closer to its degraded counterpart than to any other clean latent. We call this property "identity preservation".

\begin{table*}[t]
\centering
\resizebox{0.65\textwidth}{!}{%
\begin{tabular}{lcccccccc}
& \multicolumn{2}{c}{\textsc{vae}} 
& \multicolumn{2}{c}{CodiCodec} 
& \multicolumn{2}{c}{\textsc{dac}} 
& \multicolumn{2}{c}{Encodec} \\
\cmidrule(lr){2-3} \cmidrule(lr){4-5} \cmidrule(lr){6-7} \cmidrule(lr){8-9}
Task 
& $\cos(\theta)$ $\uparrow$ & $\Delta$\textsc{lsd} $\downarrow$
& $\cos(\theta)$ $\uparrow$ & $\Delta$\textsc{lsd} $\downarrow$
& $\cos(\theta)$ $\uparrow$ & $\Delta$\textsc{lsd} $\downarrow$
& $\cos(\theta)$ $\uparrow$ & $\Delta$\textsc{lsd} $\downarrow$ \\
\midrule
BWE
& \textbf{0.985} & \textbf{-2.086}
& \textbf{0.986} & \textbf{-0.928}
& \textbf{0.986} & \textbf{-2.286}
& \textbf{0.481} & \textbf{-1.941} \\

Denoising
& 0.627 & -0.516
& 0.596 & -0.502
& 0.596 & -0.755
& 0.464 & -0.540 \\

Declipping
& 0.337 & -0.002
& 0.290 & -0.025
& 0.189 & -0.069 
& 0.129 & +0.002  \\

Dereverberation
& 0.295 & -0.065
& 0.281 & -0.228 
& 0.323 & -0.406 
& 0.218 & +0.158 \\
\end{tabular}
}
\caption{Average cosine similarity ($\cos(\theta)$) and $\Delta$\textsc{lsd} metrics on MTD across codecs using the global mean shift transport. Lower $\Delta$\textsc{lsd} means more reduction, and hence more improvement in log-spectral distance.}
\label{tab:task_cosine_hierarchy}
\end{table*}

Turning to Figure~\ref{fig:structure_preservation_margin_orchestral} we observe that all codecs tend to exhibit margin values that are mostly positive. This suggests that bandwidth degradation is represented in the latent spaces more like a local, perhaps small, variation of the clean sample, rather than a completely different sample. As a corollary, this also means that, in large positive margin cases, modifying the bandwidth of a given sample requires a small displacement in the latent space relative to other generative tasks (e.g. generating content from noise). We argue that taking advantage of this observation may offer an opportunity for designing effective and efficient restoration systems. For example, choosing a neural codec with a large margin (or promoting it during training) ensures clear differentiation between degraded variants of an audio track and other tracks, which therefore makes the audio restoration task easier to model in the latent space. Recent work in the speech domain suggests it may be a promising avenue indeed \cite{zhang2025voicebridge}. 
Similarly, large positive margin scenarios may offer an opportunity to devise generative restoration approaches that are more efficient than simply applying a content-generation framework (e.g. from noise) to a restoration task, by taking advantage of the knowledge that the restoration transport vector required is comparatively smaller.  

Secondly, with the exception of the VAE, most latent spaces show a margin that grows with cutoff frequency. This is also consistent with stronger degradations (e.g. 4 kHz) making a bigger intra-sample difference than lighter ones (e.g. 12 kHz), which reduces the margin. It also correlates directly with reconstruction performance, where the task becomes increasingly difficult as the degradation becomes stronger.

\subsection{Restoration Task Complexity in Latent Spaces}
\label{subsec:restoration_task_complexity}

Finally, we ask whether other types of degradations would exhibit latent space structure similar to that observed in the case of bandwidth extension. Using the \textsc{mtd} dataset, we apply the same global mean-shift probe described in Section~\ref{subsec:mean_shift_local_mean} to the following restoration tasks: (i) denoising: we use three types of additive noise (white, pink, brown) across three different signal-to-noise ratios (6\,dB, 12\,dB, 18\,dB); (ii) declipping: we use hard-clipping with a signal-to-distortion ratio of 10\,dB following \cite{moliner2023solving}; (iii) dereverberation: we adopt the reverberation configuration used in generating the WSJ0-Reverb dataset~\cite{richter2023speech}. 

We evaluate the results with two metrics: the alignment of individual restoration vectors ($\cos(\theta)$), as described in Section~\ref{subsec:eval_geometry_probes}, and the resulting improvement in \textsc{lsd} ($\Delta$\textsc{lsd}) introduced by the global mean restoration vector. We report these metrics aggregated across different setups for each restoration task in Table~\ref{tab:task_cosine_hierarchy}. 
Results show that the strong behavior observed for \textsc{bwe} does not extend equally to other degradations. Denoising, declipping, and dereverberation all show lower vector alignment and smaller \textsc{lsd} gains under the same mean-shift probe.
While our simple algebraic probe does not appear to immediately reveal a representation of latent space transformations for denoising, declipping and dereverberation as crisply as what we observed for bandwidth extension, it remains an open question whether this is because the structure does not exist or because our metric is not appropriate to expose it. We leave this question for future work.



\section{Discussion}

Our results show that, for music bandwidth extension, a simple latent-space transformation can be surprisingly competitive with large generative restoration models. We do not interpret this as an end in itself, or as a replacement for learned restoration systems. Rather, it suggests that some neural codec latent spaces already contain restoration-relevant structure that larger models should be able to exploit. This raises several questions for future work.

First, how should large generative restoration models use structure that is already present in the latent representation? If a shared bandwidth-extension correction is easy to estimate, then a high-capacity model should not need to spend much of its capacity relearning this average transformation. Instead, its capacity could be directed toward the harder residual problem: sample-specific detail, perceptual realism, and codec-dependent trade-offs. 
The current training paradigms are still close to content generation frameworks, and we wonder if a variation more specific to restoration tasks may yield both more efficiency and performance? 

Second, could performance and efficiency improve if neural codec latent spaces were explicitly encouraged to represent restoration-relevant structure? Our experiments only study structure that is already present in pretrained codecs. A natural next question is whether codec training could keep clean and degraded versions of the same signal close without removing information needed for reconstruction, or encourage degradations of interest to correspond to simple latent transformations. Related latent-bridge restoration work has explored this direction~\cite{zhang2025voicebridge}, although it remains unclear how much of the reported performance comes from the representation geometry itself, and whether it improves parameter efficiency when paired with large generative models. We could not investigate the model by Zhang et al. in~\cite{zhang2025voicebridge} since it is not openly available.


Finally, how much domain structure should be injected into restoration systems? Encoding restoration-specific assumptions into the representation or model may improve short-term performance and efficiency, especially for common degradations such as bandwidth limitation. At the same time, too much task-specific structure may reduce generality across degradations, datasets, or future model classes. This tension echoes the broader trade-off between engineered domain structure and scalable general methods\footnote{\url{http://www.incompleteideas.net/IncIdeas/BitterLesson.html}}. Simple latent probes provide a lightweight way to study this trade-off before committing to larger model designs.

\section{Conclusion}

We studied audio restoration through simple deterministic probes in neural codec latent spaces. Our results show that music \textsc{bwe} can appear as a surprisingly coherent latent direction, making zero-parameter transport competitive with much larger learned restorers in several settings. This finding is not meant to replace generative restoration models, but to expose structure that such models should be able to exploit. More generally, our results suggest that probing latent geometry is a useful step before scaling restoration models: it can reveal when a degradation is already simple in the chosen representation, and when better representation learning or stronger conditional modeling is likely needed.

\bibliography{ISMIRtemplate}

\end{document}